\documentclass[showpacs,preprint,amssymb]{revtex4}
\usepackage{graphicx}
\usepackage{dcolumn}
\usepackage{bm}
\begin{document}
\title{Premium Calculation Based on Physical Principles}
\author{Amir H. Darooneh}
\affiliation{Department of Physics, Zanjan University, P.O.Box
45196-313, Zanjan, Iran.} \email{darooneh@mail.znu.ac.ir}
\date{\today}
\begin{abstract}
We consider the concept of equilibrium in economic systems from
statistical mechanics viewpoint. A new method is suggested for
computing the premium on this basis. The B\"{u}hlmann economic
premium principle is derived as a special case of our method.
\end{abstract}

\pacs{89.65.Gh, 05.20.-y} \maketitle

\section{Introduction}
Methods developed in physics are widely used for modeling and data
analysis of the financial market. This approach to quantitative
economy was adopted by economists in nineteen century and the
beginning of the twentieth century, however it was only during the
last decade that physicists turned their attention to problems of
such nature \cite{bjn}. Among all branches of physics, the
statistical mechanics appears as the most suitable context for
studying the dynamics of complex systems such as financial market
\cite{bp,ms}. There are many cases which demonstrate the power of
statistical mechanics in exploring dynamics of the financial
market. Recently Sornette and Zhou \cite{sz} show the advantage of
applying the principles of statistical mechanics to price
prediction in the US market. The study of insurance with the aid
of ideas borrowed from statistical mechanics was begun by the work
of author \cite{fd,d1,d2,d3}. In this paper we try to explain how
this approach may be used for pricing the insurance. In what
follows we first describe the equilibrium in the economic systems
from physical viewpoint then we suggest a new way for calculation
of the insurance premium. Like the economic model of Buhlmann
\cite{b1,b2} the role of market is taken into the account when
premium is assigned to a category of risks. The Esscher principle
also appears as a special case of this method.

\section{Economic Systems in Equilibrium}
The financial market is combination of large number of economic
agents which are interacting with each other through buying and
selling. We consider the behavior of one of the agents for example
an insurance company; all other agents may be regarded as its
environment. The agent exchanges money when interacts with its
environment. We suppose the financial market is a closed system
and the clearing condition is satisfied. This means, the
environment absorbs the money that the agent loses and will supply
the agent's incomes.
\begin{equation}\label{e1}
w_a+w_e=W_m=Const.
\end{equation}
The quantities $w_a$,$w_e$ and $W_m$ are the wealth of agent, its
environment and total money in the market respectively.

In a given period of time, there are many ways for the agent to
possess amounts of money as a result of random loses and incomes
in its trading. The quantity $\Gamma_a(w_a)$ represents number of
these ways. The environment has also $\Gamma_e(w_e)$ ways to
acquire amounts of money as its wealth. The state of the market is
specified by two quantities $w_a$ and$w_e$, the market has
$\Gamma_m(w_a,w_e)$ ways of reaching this specified state.
Clearly,
\begin{equation}\label{e2}
\Gamma_m(w_a,w_e)=\Gamma_a(w_a)\Gamma_e(w_e).
\end{equation}

Our common sense tells us, at any time the market chooses any one
of these ways with equal probability because no reason exists for
preferring some of them. By definition, in equilibrium state the
agent and its environment have most options for buying or selling.
Existence of any restriction disturbs the equilibrium and
decreases the number of ways that may be chosen for trading.
Mathematically this means in the equilibrium state the function
$\Gamma_m(w_a,w_e)$ should be maximized \cite{p}.
\begin{equation}\label{e3}
\frac{\partial\Gamma_m(w_a,w_e)}{\partial w_a}=0.
\end{equation}
The agent's wealth in the equilibrium state, $W_a$, is obtained by
solving the above equation. Combination of the eqs. \ref{e3} and
\ref{e2} lead us to the following equality,
\begin{widetext}
\begin{equation}\label{e4}
\left(\frac{\partial\Gamma_a(w_a)}{\partial
w_a}\right)_{w_a=W_a}\Gamma_e(W_e)+\Gamma_a(W_a)\left(\frac{\partial\Gamma_e(w_e)}{\partial
w_e}\right)_{w_e=W_e}.\frac{\partial w_e}{\partial w_a}=0.
\end{equation}
\end{widetext}
Where $W_e$ represents the equilibrium value for the environment's
wealth. By the aid of eq.\ref{e1} we have $\partial w_e /\partial
w_a=-1$, then the equilibrium condition is reduced to its new
form.
\begin{equation}\label{e5}
\left(\frac{\partial\ln\Gamma_a(w_a)}{\partial
w_a}\right)_{w_a=W_a}=\left(\frac{\partial\ln\Gamma_e(w_e)}{\partial
w_e}\right)_{w_e=W_e}.
\end{equation}
The value of parameter $\partial\ln\Gamma(w)/\partial w$ is
denoted by the symbol $\beta$, whence the condition for
equilibrium becomes,
\begin{equation}\label{e6}
\beta_a=\beta_e.
\end{equation}

The above equality is useless unless the parameter
$\beta_a(\beta_e)$ can be expressed in terms of measurable
quantities of the agent (environment). For an insurance company we
express this parameter in terms of initial wealth of company, mean
claim size and ultimate ruin probability.

\section{The Canonical Ensemble Theory in Economics}
What is the probability that the agent possesses the specified
amount $W_a^{(r)}$ when it is in equilibrium with its environment?
From basic probability theory we know it should be directly
proportional to the number of possible ways that are related to
this state of the market.
\begin{widetext}
\begin{equation}\label{e7}
Pr(W_a^{(r)})\propto\Gamma_m(W_a^{(r)},W_e^{(r)})=\Gamma_a(W_a^{(r)})\Gamma_e(,W_e^{(r)}).
\end{equation}
\end{widetext}

The environment is supposed to have much money in comparison to
the agent's wealth,
\begin{equation}\label{e8}
\frac{W_a^{(r)}}{W_m}\ll1.
\end{equation}
It is clear that $\Gamma_e(,W_e^{(r)})$ is also much larger than
$\Gamma_a(W_a^{(r)})$ hence the eq. \ref{e7} can be approximated
as,
\begin{equation}\label{e9}
Pr(W_a^{(r)})\approx\Gamma_e(W_e^{(r)})=\Gamma_e(W_m-W_a^{(r)}).
\end{equation}

We can expand the logarithm of above equation around the value .
\begin{widetext}
\begin{eqnarray}\label{e10}
\ln Pr(W_a^{(r)})&=&\ln\Gamma_e(W_m-W_a^{(r)})\nonumber \\
 &=&\ln\Gamma_e(W_m)+
 \left(\frac{\partial\ln\Gamma_e(w)}{\partial w}\right)_{w=W_m}\left(-W_a^{(r)}\right)+\cdots \nonumber \\
 &\approx&\ln\Gamma_e(W_m)-\beta W_a^{(r)}.
\end{eqnarray}
\end{widetext}

The first term in right hand side of the eq.\ref{e10} is a
constant number and the second term is nothing except of product
of parameter $\beta$ and the agent's wealth.  The eq. \ref{e8}
insures that other terms in the above expansion are small with
respect to these leading terms. By a simple algebraic manipulation
we obtain the desired result.
\begin{equation}\label{e11}
Pr(W_a^{(r)})=\frac{e^{-\beta W_a^{(r)}}}{\sum_re^{-\beta
W_a^{(r)}}}.
\end{equation}

All the accessible equilibrium states of the market make an
ensemble of possible values for the agent's wealth. The index $r$
indicates members of this ensemble. This is what the physicists
called canonical ensemble.

\section{The Insurance Pricing}
The insurance is a contract between insurer and insurant. Any
happening loss incurred on the insurant party is covered by
insurer, in return for amount of money received as premium. The
wealth of insurer at the end of time interval   is,
\begin{equation}\label{e12}
W(t)=W_0+S(t).
\end{equation}
Where $W_0$ and $S(t)$ are initial wealth and surplus of the
insurer respectively. The surplus of the insurer is the sum of all
loss events and incomes corresponding to each category of the
insurance contracts.
\begin{eqnarray}\label{e13}
S(t)&=&\sum_\alpha S_\alpha (t)\nonumber \\
 &=&\sum_\alpha (p_\alpha I_\alpha (t)-\sum_{j=1}^{N_\alpha (t)}X_j).
\end{eqnarray}
The summation goes over different categories. $I_\alpha (t)$ and
$N_\alpha (t)$ are number of the insurants and loss events for a
specified category respectively. An insurant charges the insurer
for $X_j$ in a loss event and pays a premium of $p_\alpha$ to
him/her. According to eq. \ref{e11}, the probability for acquiring
the surplus $S(t)$ by the insurer is,
\begin{equation}\label{e14}
Pr(S(t))=\frac{e^{-\beta S(t)}}{\sum e^{-\beta S(t)}}.
\end{equation}
The summation goes over all possible values for surplus. The
parameter $\beta$ is positive to ensure that extreme values for
surplus have small probability. The number of insurants in eq.
\ref{e12} is also a random variable and indicates the competition
in the market. It may be decreased due to an increment in the
premium and will be increased when the insurer reduces the prices.
In the case of constant number of insurants we obtain the result
similar to what the Buhlmann \cite{b1,b2} arrived at in his
economic model for insurance pricing.
\begin{equation}\label{e15}
Pr(Z(t))=\frac{e^{\beta Z(t)}}{\sum e^{-\beta Z(t)}}.
\end{equation}
The $Z(t)$ demonstrates the aggregate loss in the specified time
interval. The insurer naturally aims at maximizing its profit,
hence its surplus, after the insurance contract is over, should be
positive or zero at least. This condition may be expressed
mathematically only as an average form.
\begin{equation}\label{e16}
<S(t)>=\frac{\sum S(t)e^{-\beta S(t)}}{\sum e^{-\beta S(t)}}=0.
\end{equation}

When there is no correlation between financial events (loss and
gain) in different categories, the eq. \ref{e16} is reduced to
\cite{b1},
\begin{equation}\label{e17}
<S_1(t)>=\frac{\sum S_1(t)e^{-\beta S(t)}}{\sum e^{-\beta
S(t)}}=\frac{\sum S_1(t)e^{-\beta S_1(t)}}{\sum e^{-\beta
S_1(t)}}=0.
\end{equation}

In the above equation sum over all equilibrium states is
understood. The eq. \ref{e17} can be used to compute the premium
for corresponding category. For constant number of insurants it is
nothing except the Esscher formula for premium assignment to a
category of risks.
\begin{equation}\label{e18}
p=\frac{\sum Z_1(t)e^{-\beta Z_1(t)}}{\sum e^{-\beta Z_1(t)}}.
\end{equation}

So far the parameter $\beta$ has remained, however by a simple
dimensional analysis one can establish that it must be
proportional to the inverse of the contracts duration. The eqs.
\ref{e17} or \ref{e18} also shows it is related to premium. The
following equation relates ultimate ruin probability to the
premium \cite{kpw}.
\begin{equation}\label{e19}
\varepsilon=1-\sum_{k=1}^\infty
(1-\frac{p_0}{p})(\frac{p_0}{p})^kF_e^{\ast k}(W_0).
\end{equation}
Where $p_0$ is the net premium. The $F_e(x)$ is related to claim
size probability function $F(x)$,
\begin{equation}\label{e20}
F_e(x)=\frac{1}{\mu}\int_0^xF(y)dy.
\end{equation}
And $F_e^{\ast k}(W_0)$ is the k-fold convolution of $F_e(W_0)$
with itself. The parameter $\mu$ represents the mean claim size.
The ruin probability, initial wealth and mean claim size tune the
financial work of insurer and are concealed in the   parameter.
Combination of the eqs. \ref{e17} and \ref{e19} shows this fact.
\begin{figure}
\includegraphics[width=8.5cm,height=5.5cm]{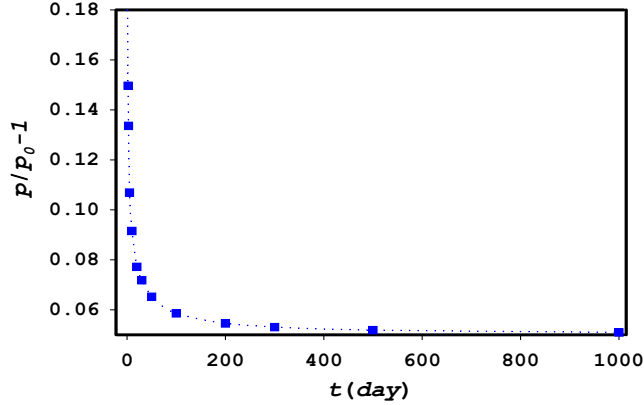}
\caption{\label{f1} The loading parameter versus the contract's
duration for large $\beta$ parameter.}
\end{figure}
\begin{figure}
\includegraphics[width=8.5cm,height=5.5cm]{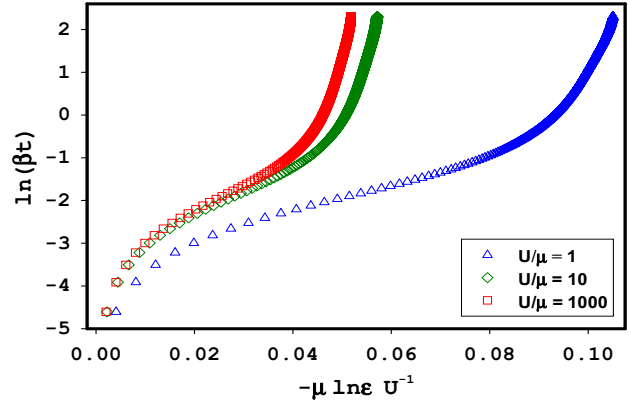}
\caption{\label{f2} Dependence of the $\beta$ parameter on the
ruin probability, initial wealth and mean claim size. }
\end{figure}
\begin{figure}
\includegraphics[width=8.5cm,height=5.5cm]{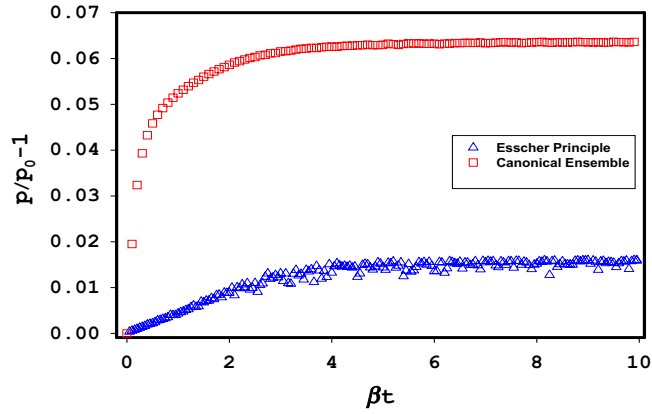}
\caption{\label{f3} The loading parameter versus the $\beta t$
parameter. The squares display the results of the canonical
ensemble theory and triangles correspond to Esscher principle.}
\end{figure}

Below, the simulation results for a special case of car insurance
are presented for demonstrating some aspects of our method. For
simplicity we assume that all cars are the same and pay the same
premium, the loss events have exponential distribution for time
interval between their occurrence and their size. The necessary
data are adopted from reports of Iran Central Insurance Company
\cite{r}. Fig. \ref{f1} demonstrates dependence of loading
parameter on the contract's time. It is what we expected in real
cases. As we already mentioned, if we specify the ruin
probability, initial wealth and mean claim size the $\beta$
parameter will be determined. Fig. \ref{f2} displays the relation
between these parameters. The initial wealth is measured in terms
of the mean claim size in order to get ride of dependence of the
results on the monetary unit. The variation in number of the
insurants influences the premium. Fig.\ref{f3} is plotted to show
this effect, the value of the premium which is obtained from
Esscher formula is less than that we found in our method.

\section{Conclusion}
Statistical mechanics are pervasively used for studying the
financial behavior of an economic agent in the market. The concept
of equilibrium is revised in this respect. We prescribe it for the
insurance subject and suggest a new way for computing the premium.
The effects of interaction between insurer and the market are
included in this new method. Some consequences of B\"{u}hlmann
economic principle are recovered. This work has been supported by
the Zanjan university research program on Non-Life Insurance
Pricing No: 8243.

\bibliographystyle{}

\end{document}